\documentclass[aps,pra,twocolumn,superscriptaddress,10pt]{revtex4-1}
\usepackage{amsmath,amssymb,graphicx}

\usepackage{color}

\newcommand{\bitem}{\begin{itemize}}
\newcommand{\fitem}{\end{itemize}}
\newcommand{\beq}{\begin{equation}}
\newcommand{\eeq}{\end{equation}}
\newcommand{\beqa}{\begin{eqnarray}}
\newcommand{\eeqa}{\end{eqnarray}}

\begin{document}

\title{Entanglement detection via atomic deflection}

\author{C. E. M\'aximo} \affiliation{Instituto de F\'{i}sica de S\~ao Carlos, Universidade de S\~ao Paulo, 13560-970 S\~ao Carlos, SP, Brazil}

\author{R. Bachelard} \affiliation{Instituto de F\'{i}sica de S\~ao Carlos, Universidade de S\~ao Paulo, 13560-970 S\~ao Carlos, SP, Brazil}
\affiliation{Departamento de F\'{\i}sica, Universidade Federal de S\~{a}o Carlos, Rod. Washington Lu\'{\i}s, km 235 - SP-310, 13565-905 S\~{a}o Carlos, SP, Brazil}

\author{G. D. de Moraes Neto} \affiliation{Instituto de F\'{i}sica, Universidade Federal de Uberl\^{a}ndia, Uberl\^{a}ndia, Minas Gerais 38400-902, Brazil}  

\author{M. H. Y. Moussa} \affiliation{Instituto de F\'{i}sica de S\~ao Carlos, Universidade de S\~ao Paulo, 13560-970 S\~ao Carlos, SP, Brazil}

\begin{abstract}
We report on criteria to detect entanglement between the light modes of two crossed optical cavities by analyzing the transverse deflection patterns of an atomic beam. The photon exchange between the modes and the atoms occurs around the overlapping nodes of associated standing waves, which generates the two-dimensional (2D) version of the Optical Stern-Gerlach  (OSG) effect. In this optical cross-cavity setup, we show that the discrete signatures of the field states, left in the momentum distribution of the deflected atoms, may reveal entanglement for a certain class of two-mode states. For a single photon, we present the possibility of quantifying entanglement by the rotation of the momentum distribution. For a larger number of photons, we demonstrate that quantum interference precludes the population of specific momentum states revealing maximum entanglement between the light modes.

\bigskip

{\fontsize{8}{8}\selectfont  \textbf{OCIS codes:} (020.5580)  Quantum electrodynamics; (140.4780) Optical resonators; (020.1335) Atom optics}
\end{abstract}

\maketitle

\section{Introduction}

Light is the cornerstone of various techniques to probe the state of matter, be they ordered, such as crystalline structures for which Bragg's law will be used~\cite{bragg1913}, or disordered, for which speckle analysis~\cite{butters1971} or diffusive wave spectroscopy~\cite{maret1987} can still reveal precious information. One can even probe the reciprocal space of the system (i.e., the momentum space), provided the atoms have well-defined momenta. This was for example achieved in Bose-Einstein condensates~\cite{stenger1999}.

Reciprocally, the mechanical effects of light on the atoms can serve as an important tool to extract information on radiation fields. In the context of ultracold matter, the momentum distribution is accessed through the time-of-flight technique, where the recoil acquired by the particles provide information on the light modes populated. For instance, the scattering properties of Bose-Einstein condensates occur in specific directions, which in turn corresponds to discrete momentum states~\cite{inouye1999}.

The difficulty in measuring or extracting information from a three-dimensional pattern leads to considering effectively low-dimensional configuration, by using, for example, optical cavities. The advantage is twofold: The number of light modes available is quite precisely controlled~\cite{kleppner1981}, and the requirement of well-defined momentum of the atoms need to be achieved only along the axis of the cavity. Then, even the quantum nature of the light may be probed. Indeed in cavity quantum electrodynamics, the strong coupling between a single atom and a field mode imprints signatures of the discrete nature of the field in the momentum distribution of deflected atoms~\cite{meystre1989}. As a result, some works were dedicated to reconstructing the photon statistics of the radiation field by performing demolition \cite{schleich1992} and non-demolition~\cite{zoller1991} measurements on the atoms. The phases of the field can also be computed, up to a global one~\cite{herkommer1994}, by collimating the atomic beam into a node of a standing wave. This configuration creates a quantum symmetrical beam splitter, analog to the original Stern-Gerlach experiment with a magnetic field, and is now referred to as ``Optical Stern-Gerlach`` (OSG)~\cite{mlynek1996}.

The works of the last paragraph have in common a single mode treatment, nevertheless, a multi-mode configuration brings up the possibility to obtain information on a crucial ingredient of quantum mechanics: entanglement. A common protocol introduced to measure entanglement consists of the Wigner function reconstruction of more-than-one-mode states. Among these, one can cite the atomic deflection by two successive cavity modes operating in the Bragg regime~\cite{zubairy2004}, the atomic deflection by two modes within the same cavity~\cite{zubairy2005} and the multi-mode-state measurement of traveling fields~\cite{song2014}. In this paper, we report on a protocol to detect entanglement without relying on the reconstruction of the full state of the system. We present simple criteria to detect entanglement between two modes, in an OSG setup~\cite{maximo2014}, by identifying simple features in 2D atomic deflection patterns. 

In typical cavities quantum electrodynamics experiments, a configuration which achieved one of the best control of atom-photon interaction~\cite{haroche2001}, only the internal degrees of freedom of the atoms are used to probe a field mode state. The introduction of the external degrees of freedom in the analysis, thanks to the orthogonal crossed cavities, makes a single atom a sufficient meter to obtain information about two-mode states. It thus opens the possibility to detect entanglement without measurements other than the 2D atomic momentum distribution. In particular, we demonstrate that one-photon states offer an easy reading since the concurrence can be directly read off from the rotation of the deflection pattern. Finally, NOON states and other highly entangled states are shown to be characterized by the disappearance of specific momenta states due to quantum interference. 

In section II, we present a description of the cross-cavity setup, exhibiting a formal analytical solution for the 2D atomic momentum distribution. The  signatures of entanglement between the two cavity modes are investigated in section III. Finally, we draw our conclusions in section IV.

\section{The cross-cavity setup}

\begin{figure}[h!]
\begin{center}
\includegraphics[scale=0.54]{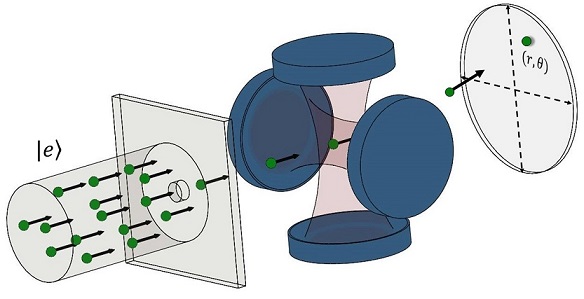}
\end{center}
\caption{Cross-cavity OSG setup: A collimated atomic beam interacts with two cavity modes before traveling as free particles until the screen, where the two-dimensional transverse deflection pattern is measured. The spatial distribution of the deflected atoms, by means of time-of-flight technique, reveals features on the cavity modes through the momentum exchanged between the atoms and the photons.}\label{fig:Setup}
\end{figure}

The cross-cavity OSG setup is depicted in Fig.\ref{fig:Setup}. A beam of two-level atoms traveling along the $z$ axis, with transition frequency $\omega_{0}$, crosses two identical optical cavities which are along the  $x$ and $y$ axes. A single mode from each cavity, with frequency $\omega=kc=2\pi c/\lambda$ and Rabi frequency $g$, interacts simultaneously and resonantly with the atoms. The cavities present orthogonal electric fields so each imprints momentum to the atoms in a different direction. Since the longitudinal kinetic energy of the atoms is much larger than the typical atom-field coupling, the longitudinal motion is essentially unaffected by the interaction and is treated classically. The change in the atomic transverse kinetic energy $E_\text{kin}^\perp$ is also neglected under the Raman-Nath regime, when $E_\text{kin}^\perp\ll\sqrt{n}\hbar g$, with $n$ the number of photons in the system. Nevertheless, the change in momentum, which yields a direct signature of its interaction with the two optical modes, is treated quantum-mechanically.

The Stern-Gerlach regime is achieved by introducing a narrow circular slit in front of the cavities array to collimate transversally the atomic beam in a subwavelength region, around the overlapping nodes of the standing-waves, similarly to the single cavity case~\cite{herkommer1994}. This allows to linearize the cavity fields ($\hat{\mathbf{y}}\sin kx \approx \hat{\mathbf{y}}kx$ and $\hat{\mathbf{x}} \sin ky \approx \hat{\mathbf{x}}ky$) and to describe the interaction between the atoms and the photons by the following Hamiltonian:
\begin{equation}
H=-\hbar g kr\left(\cos\theta\text{ }a+\sin\theta\text{ }b\right)\sigma_{+}+h.c.,\label{h}
\end{equation}
where polar coordinates have been used. The operators $a$ ($a^{\dagger}$) and $b$ ($b^{\dagger}$) stand for the annihilation (creation) operators of the cavity modes with optical axis in the $x$ and $y$ directions, respectively, while $\sigma_{+}=\left\vert e\right\rangle \left\langle g\right\vert $ and $\sigma_{-}=\left\vert g\right\rangle \left\langle e\right\vert $ describe the raising and lowering operators for the atomic transition between the ground and excited states.

Concerning to initial states, the cavity modes are initially prepared in an arbitrary pure state 
\begin{equation}
\left\vert\psi_{field}\right\rangle=\sum_{m.n=0}^{\infty}\mathcal{C}_{m,n}\left\vert m,n\right\rangle _{ab},
\end{equation}
with $\mathcal{C}_{m,n}$ the two-mode probability amplitudes. The two-level atoms are prepared in the excited state $\left\vert e\right\rangle $,
with a de Broglie atomic wave packet described by 
\begin{equation}
\left\vert \psi_{atom}\right\rangle = \left\vert e\right\rangle \int\nolimits _{0}^{\infty}\int\nolimits _{0}^{2\pi}drd\theta r f(r,\theta)\left\vert r,\theta\right\rangle, \label{istate}
\end{equation}
where $\left\vert f(r,\theta)\right\vert ^{2}$ corresponds to the spatial distribution of the atoms right after the circular slit. We here consider that the system is closed since the interaction time $\tau$ between the cavity and the atoms is short enough to neglect spontaneous emission. This regime can be achieved, for example, using atoms excited to circular Rydberg levels, as they present a strong coupling to the cavity and long decay times~\cite{raimond2015}. Atoms which do not populate the desired Rydberg states will end at center of the spatial momentum pattern, suffering only slight deflection from diffraction. Furthermore, we note that for times over which spontaneous emission kicks in, information on the photon pattern can still be obtained from the diffuse momentum distribution, as was shown by~\cite{schleich1994}. 

The interaction between the light and the atoms leaves its signatures in the momentum pattern of particles, which can be recovered by the time-of-flight technique. The latter technique has been used in single cavity OSG setup~\cite{herkommer1994}, but is also common with ultracold matter~\cite{inouye571}: After the interaction within the cavity volume, the atom-atom interactions are negligible so they travel as free particles. At later times as the initial spatial spread has become negligible, the spatial distribution is an accurate description of the transverse momentum distribution. Therefore, all the analysis can be presented in terms of the dimensionless atomic momentum $\wp=p/\hbar k$ and the momentum angle $\phi$, which satisfy the relation $(p_{x},p_{y})=p(\sin\phi,\cos\phi)$. 
\begin{figure*}
\begin{center}
\includegraphics[scale=0.59]{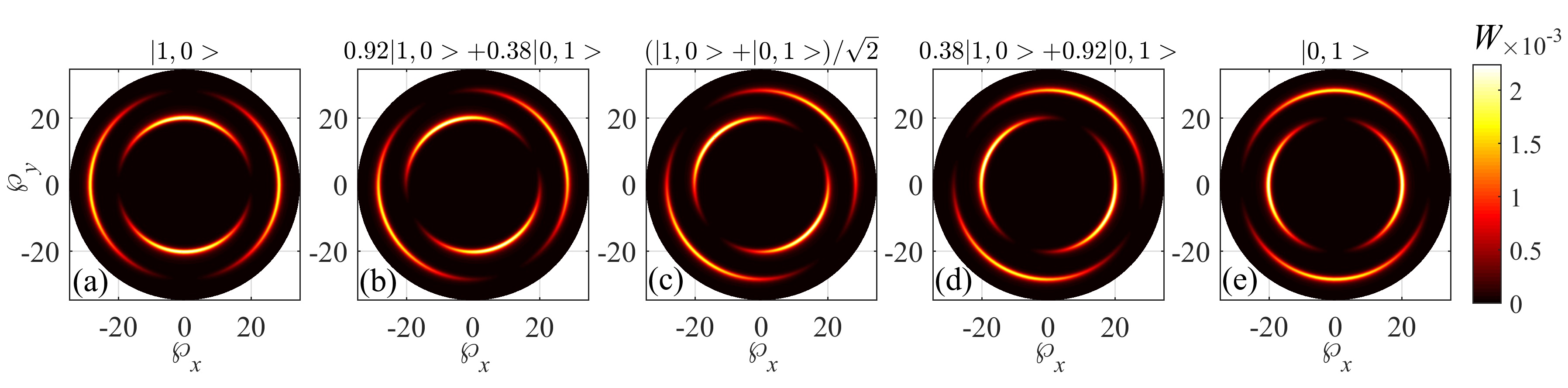}
\end{center}
\caption{Atomic deflection patterns for five configurations of the one-photon states $\sin\alpha\left\vert 0,1\right\rangle+\cos\alpha\left\vert 1,0\right\rangle$. The second to fourth panels are entangled states, the leftmost and rightmost ones correspond to separable states. Other values of $\alpha$ (between $\pi/2$ and $2\pi$) present the same features, due to the rotational symmetries of the system. Simulations realized for $\Lambda=20$, $k\Delta r=0.1$, and $\alpha=0,\pi/8, \pi/4, 3\pi/8, \pi/2$ for panel (a-e), respectively.}\label{fig:0110} 
\end{figure*}

In the present work, the 2D transverse momentum profile is directly computed from the density matrix of the evolved system, tracing over the Fock states and the internal degrees of the atoms. The resulting momentum distribution reads (see Appendix.)
\begin{widetext} 
\begin{align}
W\left(\wp,\phi,\Lambda\right)=\frac{1}{2}\sum_{N=1}^{\infty}\sum_{n=1}^{N}\left[\left|\sum_{m=1}^{N}\mathcal{C}_{m-1,N-m}\mathcal{F}_{m,n}^{+\left(N\right)}(\wp,\phi,\Lambda)\right|^{2} + \left|\sum_{m=1}^{N}\mathcal{C}_{m-1,N-m}\mathcal{F}_{m,n}^{-(N)}(\wp,\phi,\Lambda)\right|^{2}\right], \label{w}
\end{align}
where $\Lambda=g \tau$ stands for the atom-field interaction parameter. The the two-dimensional Fourier transform has been defined as 
\begin{align}
\mathcal{F}_{m,n}^{\pm(N)}(\wp,\phi,\Lambda)=\int_{0}^{\infty}\int_{0}^{2\pi}\frac{kr}{2\pi}drd\theta f\left(r,\theta\right)D_{m-1,n-1}^{(N-1)}\left(\theta\right)\operatorname*{e}\nolimits ^{-ikr\left[\wp\cos(\theta-\phi)\mp\sqrt{n}\Lambda\right]}, \label{f}
\end{align}
\end{widetext}
with $D_{m,n}^{(N)}(\theta)$ the diagonalization coefficients, which is explicit precisely in the appendix. In Eq.\eqref{f}, the  functions $\mathcal{F^{\pm}}$  depend on  integer numbers: The number of photons in the modes on the system $N=N_\nu+1$, where $N_\nu$ is the number of photons only in the cavities and the index $n$, which appears in particular in the exponential argument $-ikr\left[\wp\cos(\theta-\phi)\mp\sqrt{n}\Lambda\right]$, counts the almost-discrete momentum states. Indeed the momentum distribution is concentrated around $\wp=\sqrt{n}\Lambda$. The $\pm$ signs highlight the existence of a more general beam splitter in 2D, a generalization of the OSG effect. We here choose an exponential density
\begin{equation}
f(r,\theta)=\frac{1}{\sqrt{2\pi}\Delta r}\exp(-r/2\Delta r)
\end{equation}
that yields non-trivial analytical expressions of the Fourier integrals in Eq.\eqref{f} (see Appendix). Considering the exponential spatial distribution of the atoms instead of, for example, a Gaussian profile allows us to obtain an analytical expression for the atomic momentum distribution, and thus reduce the computational cost of the routine. Yet arbitrary distributions can be addressed by our procedure, without change to the underlying physics: The momentum distribution peaks of $W(\wp,\phi,\Lambda )$ is always centered around the same values $\wp=\sqrt{n}\Lambda $ independently of the shape of $f(r,\theta)$, as shown by  Eq.~\ref{f}.
\begin{figure}
\begin{center}
\includegraphics[scale=0.50]{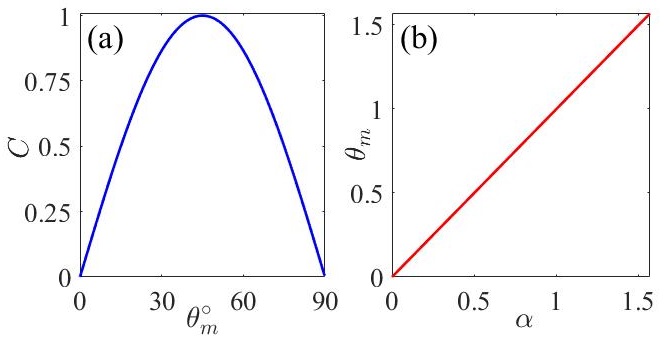}
\end{center}
\caption{(a) Concurrence $C$ as a function of the angle $\theta_{m}$, which corresponds to the measured maximum of the momentum state ($\wp=\sqrt{2}\Lambda$).(b) The correspondence between the measured angle $\theta_{m}$ and $\alpha$ is shown. Same parameters as in Fig.\ref{fig:0110}.}\label{fig:conctheta}
\end{figure}

\section{Entanglement detection}
\begin{figure*}
\begin{center}
\includegraphics[scale=0.57]{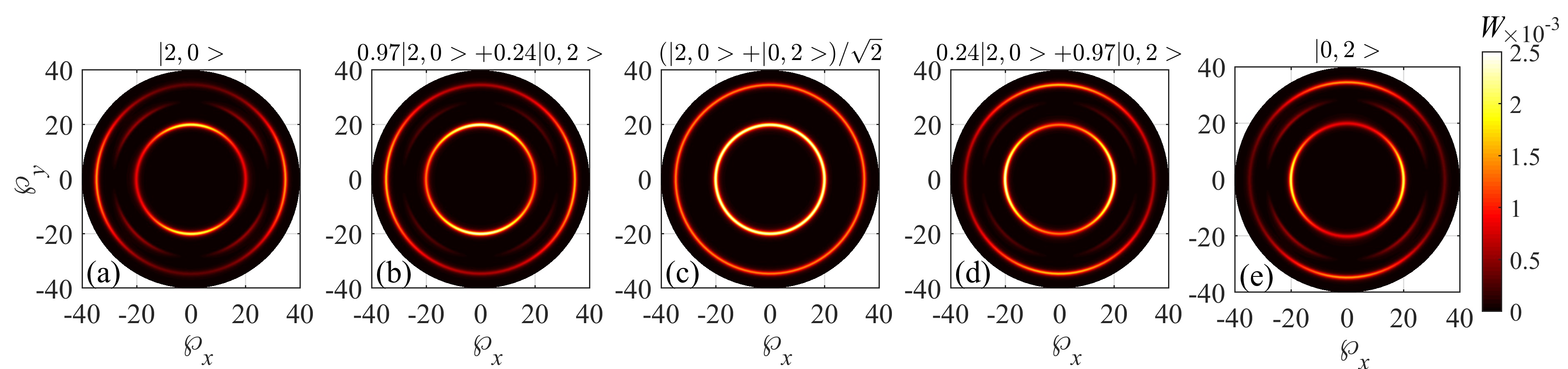}
\end{center}
\caption{Atomic deflection pattern for the $\cos\alpha|2,0\rangle+\sin\alpha|0,2\rangle$ cavities state, for $\alpha=0$, $\pi/13$, $\pi/4$, $11\pi/26$ and $\pi/2$. Same parameters as Fig.\ref{fig:0110}.}\label{fig:2002}
\end{figure*}

 Let us now discuss how the 2D deflection patterns can be an entanglement signature. At first, we focus on the states of the form $\left\vert\psi\right\rangle=\sin\alpha\left\vert 0,1 \right\rangle+\cos\alpha\left\vert 1,0 \right\rangle$: these are the simplest one-cavity-photon states ($N_\nu=1$) that may exhibit entanglement. A series of deflection patterns is presented in Fig.\ref{fig:0110}, where $\alpha$ is tuned from $0$ to $\pi/2$. After the interaction, the momentum distribution $W$ is concentrated around $\wp=\sqrt{n}\Lambda$, where $n=1,2$ since it is limited by the total excitation $N=2$. One then observes that the atomic pattern rotates precisely with $\alpha$, which measures the balance between the two modes. When $\alpha$ reaches $\pi/4\mod(\pi/2)$, the point of maximal entanglement, the recoil distribution is now concentrated off the cavity axes. Therefore, the reference angle $\theta _{m}$ of the momentum pattern rotation, which can be defined as the angle satisfying the condition $\max \{ W\left(\wp=\sqrt{2}\Lambda,\phi=\theta _{m},\Lambda\right) \}$, is a direct measurement of the entanglement since it can be directly mapped to the concurrence of the one-photon pure state $C=\left|\sin2\alpha\right|=\left|\sin2\theta_m \right|$. (see Fig.\ref{fig:conctheta})~\cite{horodecki2009}. In other words, the closer  $\theta_{m}$ is to $\pi/4\mod(\pi/2)$, the more entangled the two cavity modes are. Note that a simple superposition of the deflection pattern obtained from states $\left\vert 0,1 \right\rangle$ and $\left\vert 1,0 \right\rangle$  could not produce the observed rotating pattern, so the latter is the result of quantum interference.

Let us then move to higher photon number states. Among these the NOON states $\left\vert\psi\right\rangle=(\left\vert N_\nu,0\right\rangle+\left\vert 0,N_\nu\right\rangle)/\sqrt{2}$ have attracted a growing interest due to their possibility to beat the standard quantum limit~\cite{sanders1989,boto2000,wolfgramm2012}, with strong implications for metrology. In the cross-cavity setup, these correspond to states where each cavity has a $50\%$ chance to have the $N_\nu$ photons when the other has none. Through the OSG, the NOON states leave a non-intuitive characteristic signature on the deflection pattern: For $N_\nu=2\mod(4)$, no atomic population is observed at $\wp=\sqrt{N_\nu/2+1}\Lambda$, whatever is the angle $\phi$. This is well illustrated in Fig.\ref{fig:2002} where the deflection pattern of the $\sin\alpha\left\vert 0,2\right\rangle+\cos\alpha\left\vert 2,0\right\rangle$ state is presented for different angles $\alpha$ between $0$ and $\pi/2$. In this case, while the $\wp=\sqrt{2}\Lambda$ state has atomic population for the $\left\vert 2,0\right\rangle$ and $\left\vert 0,2\right\rangle$ states, it progressively disappears as $\alpha$ approaches $\pi/4$, which is precisely the point of maximum entanglement. Thus the full disappearance of the atomic population in this state acts as a signature of maximum entanglement in the system. Note that a closer inspection actually reveals a very low population, which probably originates in the tails of the neighboring momentum states. Higher values of the coupling parameter $\Lambda$ lead to better-resolved momentum states and even lesser population in the disappearing momentum state~\cite{herkommer1994,maximo2014}.
\begin{figure}[h]
\begin{center}
\includegraphics[scale=0.7]{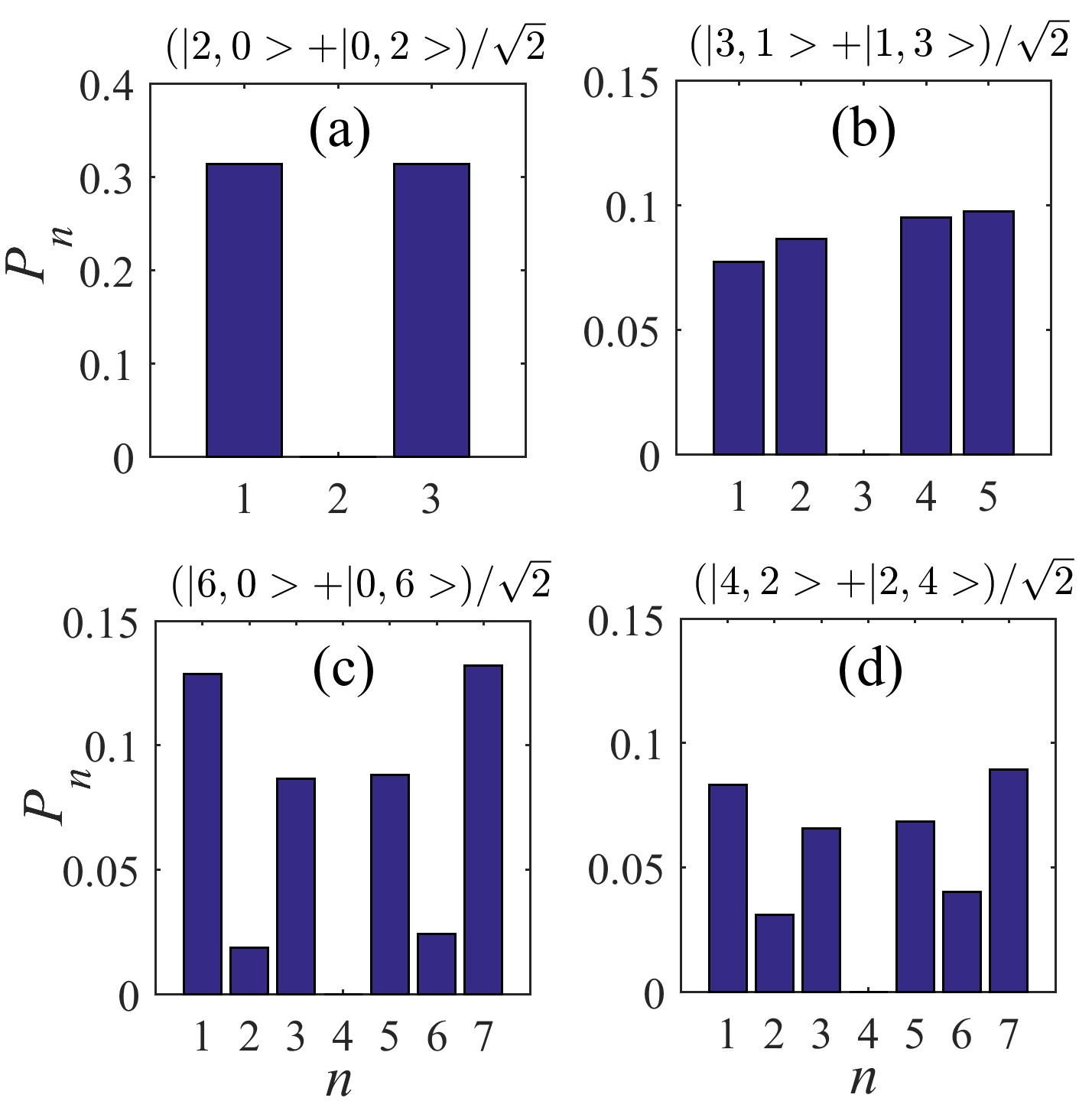}
\end{center}
\caption{Population of the atomic momentum states $\wp=\sqrt{n}\Lambda$, with $n=0,...N_\nu$ for different entangled states. The absence of population of a specific momentum state is a signature of maximum entanglement between the cavity modes (see main text). Simulations realized for $\Lambda=100$, $k\Delta r=0.1$.\label{fig:dip}}
\end{figure}

The destructive quantum interference emerging at the atomic momentum distribution is not unique to NOON states. This feature appears to be valid for an infinite class of maximally entangled states, namely the $(\left\vert j,j+4q-2\right\rangle +\left\vert j+4q-2,j\right\rangle)/\sqrt{2}$ states, with $j=0,1,...$ and $q=1,2,...$  the indexes range. The number of cavity photons of these states is always even ($N_\nu = 2(j+2m-1)$), with the NOON states corresponding to $j=0$.

Let us have a closer look at these states, calculating the population of deflected atom in each momentum state $\wp=\sqrt{n}\Lambda$. As discussed in Refs.~\cite{herkommer1994,maximo2014}, for large coupling parameters $\Lambda \gg 1$, the Fourier transforms \eqref{f} are very close to Delta functions centered around the momentum state $\wp=\sqrt{n}\Lambda$. It is therefore reasonable to define the discrete measurable probabilities
\begin{equation}
P_{n}\approx\Lambda\sqrt{n}\int_{0}^{2\pi}W\left(\wp=\Lambda\sqrt{n},\phi,\Lambda\right)d\phi.\label{P}
\end{equation}

As presented in Fig.\ref{fig:dip} for photon numbers $N_\nu=2$ (panel (a)), $N_\nu=4$ (panel (b)) and $N_\nu=6$ (both panels (c) and (d)), the population of momentum state $\wp=\sqrt{N_\nu/2+1}\Lambda$ is clearly missing (which corresponds resp. to $\sqrt{2}\Lambda$, $\sqrt{3}\Lambda$ and $\sqrt{4}\Lambda$). Panels (c) and (d) highlight the existence of different entangled states exhibiting the same unpopulated momentum state for $N_\nu\geq4$.  Table~\ref{tab} summarizes these results by associating a missing momentum state to the corresponding entangled states.
\begin{table}[h!]
\begin{center}
\begin{tabular}{ | c | c | c | c | c | }
 \hline
 Unpopulated $\wp$ & $j=0$ (NOON) & $j=1$ & $j=2$ & $\cdots$  \\ 
 \hline
 $\sqrt{2}\Lambda$ & $\left\vert 0,2\right\rangle$ & - & - & $\cdots$ \\  
 \hline
 $\sqrt{3}\Lambda$ & - &  $\left\vert 1,3\right\rangle$ & -& $\cdots$ \\
 \hline
 $\sqrt{4}\Lambda$ & $\left\vert 0,6\right\rangle$ & - & $\left\vert 2,4\right\rangle$ & $\cdots$  \\ 
 \hline
 $\sqrt{5}\Lambda$ & - &  $\left\vert 1,7\right\rangle$ & -& $\cdots$ \\
 \hline
 $\sqrt{6}\Lambda$ & $\left\vert 0,10\right\rangle$ & - & $\left\vert 2,8\right\rangle$ & $\cdots$  \\ 
 \hline
 $\vdots$ & $\vdots$ & $\vdots$ & $\vdots$ & $\ddots$  \\ 
 \hline
\end{tabular}
\end{center}
\caption{Table of entangled states and the corresponding unpopulated momentum state. For simplicity, we display only one ket.\label{tab}}
\end{table}

Entanglement in higher photon number states, where the recoil distribution spread over many momentum states, may be challenging as the distance between two successive momentum peaks decreases ($\sqrt{n+1}\Lambda-\sqrt{n}\Lambda\sim \Lambda/\sqrt{n}$) and the interaction parameter is finite. Considering a hot beam of Rb atoms with typical speeds $\sim 500$m/s ~\cite{haroche2001,raimond2015}, in the microwave regime ($k \sim 2 \pi \times 10^{2}$m$^{-1}$) an angular separation $\sim \sqrt{n}\Lambda \times 10^{-10}$ is obtained. In the strong coupling regime, the coupling parameter can be set to $\Lambda = g\tau \sim 100$, which leads angular separation $10^{-8}$rad for single photons ($n=1$). For a de Broglie atomic wavelength of $\lambda_{dB}\sim 10^{-11}$m and a pinhole of radius $\Delta r=30$mm (equivalent to the value $k \Delta r=0.1$ for the figures of the manuscript), the diffraction angle is $\lambda_{dB}/\pi \Delta r \sim 10^{-10}$rad, which is two orders of magnitude smaller than the above deflection angle.  Therefore  diffraction can be neglected in the strong-coupling regime, and even more for larger photon numbers ($n>1$).  There numbers however imply that at a distance of about one meter from the cavity, the 2D deflection pattern has to be resolved at the nanometer scale. Based on the unidimensional OSG experiment~\cite{mlynek1996}, such experiment will require at least hundreds of thousands of repetitions to measure precisely the  angle $\theta_m$.

 \section{Conclusions}

As a conclusion, we have shown how a cross-cavity OSG setup provides simple criteria to entanglement detection without reconstructing the whole state of the system. The rotation of the atomic pattern and the missing momentum states can be used to detect entanglement of two-mode states.  For one photon states, the former criterion is a first step toward obtaining a more general entanglement quantifier when full state reconstruction is not possible. For higher photon numbers, the absence of specific momentum states acts as a detection of maximal entanglement. A natural extension of this work is to generalize these results to broader classes of nonclassical states, within the framework of mixed states.  Finally, it is worth considering an open systems approach, by taking into account spontaneous emission and decoherence effects. 

\begin{flushleft}
\textbf{\Large{}{}Acknowledgements} 
\par\end{flushleft}

The authors acknowledge the support from FAPESP(Grant Nos. 2014/19459-6 and 2014/01491-0), CNPQ and CAPES, Brazilian
agencies.

\bibliographystyle{unsrt}
\bibliography{ref}

\onecolumngrid
\section*{Appendix: Analytical solution of Eq.~\ref{f}}

In order to derive the 2D momentum distribution of the main text, we apply the canonical Bogoliubov transformation $c=\cos\theta a+\sin\theta b$ and $d=-\sin\theta a+\cos\theta b$, that takes the original field operators ($a$ and $b$) to the transformed ones ($c$ and $d$). Such change preserves the total photons number $\hat{N}_\nu=a^{\dagger}a+b^{\dagger}b=c^{\dagger}c+d^{\dagger}d$ and enables us to simplify the Hamiltonian from Eq.\eqref{h} to the form 
\begin{equation}
\mathcal{H}=-\hbar g kr\left(\sigma_{+}c+\sigma_{-}c^{\dagger}\right).\label{h1}
\end{equation}

Once the original and transformed operators share the same vacuum state, the Fock product states in both spaces $\left\{\left\vert m,n\right\rangle_{ab}\right\}$  and $\left\{ \left\vert m,n\right\rangle _{cd}\right\}$ are related to each other as  
\begin{equation}
\left\vert m,N_\nu-m\right\rangle_{ab}={\textstyle \sum\nolimits_{n=0}^{N_\nu}}D_{m,n}^{(N_\nu)}(\theta)\left\vert n,N_\nu-n\right\rangle _{cd},\label{expansion}
\end{equation}
where $N_\nu$ indicates the total photons number of both modes, while the expansion coefficients are given by 
\begin{equation}
D_{m,n}^{(N_\nu)}(\theta)=\sum_{q=\max(0,m+n-N_\nu)}^{\min(n,m)}\overline{D}_{m,n,q}^{(N_\nu)}\left(\cos\theta\right)^{N_\nu-m-n+2q}\left(\sin\theta\right)^{m+n-2q},
\end{equation}
with amplitude 
\begin{equation}
 \overline{D}_{m,n,q}^{(N_\nu)}=\frac{(-1)^{m-q}\sqrt{m!n!\left(N_\nu-m\right)!\left(N_\nu-n\right)!}}{q!(m-q)!\left(n-q\right)!(N_\nu-m-n+q)!}.
\end{equation}

With Hamiltonian \eqref{h1} and relation \eqref{expansion} in hands, we then turn our attention to the initial state of the system. Before entering the cavities, the two-level atoms (ground $\left\vert g \right\rangle$ and excited $\left\vert e \right\rangle$ states) are prepared, in a Ramsey zone, in the superposition state $c_{g}\left\vert g\right\rangle +c_{e}\left\vert e\right\rangle$, such that the de Broglie atomic wave packet crossing the cross-cavity array is given by 
\begin{equation}
\left\vert \psi_{atom}\right\rangle ={\textstyle \int\nolimits_{0}^{\infty}}{\textstyle \int\nolimits_{0}^{2\pi}}drd\theta rf(r,\theta)\left\vert r,\theta\right\rangle \left(c_{g}\left\vert g\right\rangle +c_{e}\left\vert e\right\rangle \right),
 \end{equation}
where $\left\vert f(r,\theta)\right\vert ^{2}$ accounts for the spatial distribution of the atoms, as determined by the circular slit. Regarding the cavity modes, we assume that they are initially prepared in the arbitrary pure state 
\begin{equation}
\left\vert \psi_{field}\right\rangle =\sum_{m.n=0}^{\infty}\mathcal{C}_{m,n}\left\vert m,n\right\rangle _{ab}.
 \end{equation}
  Then, the initial state of the whole system in the basis of $c$ and $d$ operators can be written as 
\begin{eqnarray}
 \left\vert \Psi\left(r,\theta,0\right)\right\rangle &=&rf\left(r,\theta\right)c_{g}\sum_{N=0}^{\infty}\sum_{m=0}^{N}\mathcal{C}_{m,N-m}D_{m,0}^{(N)}\left(\theta\right)\left\vert g,0,N\right\rangle _{cd}+\frac{rf\left(r,\theta\right)}{\sqrt{2}}\sum_{N=1}^{\infty}\sum_{n=1}^{N}\left\vert N-n\right\rangle _{d}\\ &&
	\times\left[\left(c_{g}\sum_{m=0}^{N}\mathcal{C}_{m,N-m}D_{m,n}^{(N)}\left(\theta\right)+c_{e}\sum_{m=1}^{N}\mathcal{C}_{m-1,N-m}D_{m-1,n-1}^{(N-1)}\left(\theta\right)\right)\left\vert +,n\right\rangle _{c}\right.\\ &&
	+\left.\left(c_{g}\sum_{m=0}^{N}\mathcal{C}_{m,N-m}D_{m,n}^{(N)}\left(\theta\right)-c_{e}\sum_{m=1}^{N}\mathcal{C}_{m-1,N-m}D_{m-1,n-1}^{(N-1)}\left(\theta\right)\right)\left\vert -,n\right\rangle _{c}\right]\label{psi0}
\end{eqnarray}
where we have taken advantage of the eigenstates of $\mathcal{H}$: $\left\vert \pm,n\right\rangle _{c}=\left(\left\vert g,n\right\rangle _{c}\pm\left\vert e,n-1\right\rangle _{c}\right)/\sqrt{2} (n\geq1)$. Observe that, by using the dressed states $\left\vert\pm ,n\right\rangle _c$, from Eq.\eqref{psi0} onwards, we conveniently reindex the notation to the total excitation number of the system $N=N_\nu+N_e$, with $N_e$ the excitation number of the atom. 

Once we diagonalized the initial state, the total system evolution is trivial, namely 
\begin{equation}	
e^{-i\mathcal{H}\tau/\hbar}\left\vert \pm,n\right\rangle _{c}=e^{\pm i\Lambda kr}\left\vert \pm,n\right\rangle _{c},
\end{equation}
where the spatial distribution of the atoms at $t=\tau$ can be computed via the probability distribution in momentum space, by using the time-of-flight technique. For this purpose, we first calculate the evolved state of the system in momentum representation, and then the system density matrix by tracing over the Fock states and the internal degrees of the atoms. Therefore we obtain the atomic momentum probability density
\begin{eqnarray}
W\left(\wp,\phi,\Lambda\right)	&=&\left|c_{g}\right|^{2}\left|\sum_{N=0}^{\infty}\sum_{m=0}^{N}\mathcal{C}_{m,N-m}\mathcal{F}_{m,0}^{g(N)}(\wp,\phi)\right|^{2}\\ &&
	+\frac{1}{2}\sum_{N=1}^{\infty}\sum_{n=1}^{N}\left|c_{g}\sum_{m=0}^{N}\mathcal{C}_{m,N-m}\mathcal{F}_{m,n}^{+g(N)}(\wp,\phi,\Lambda)+c_{e}\sum_{m=1}^{N}\mathcal{C}_{m-1,N-m}\mathcal{F}_{m,n}^{+e\left(N\right)}(\wp,\phi,\Lambda)\right|^{2}\\ &&
	+\frac{1}{2}\sum_{N=1}^{\infty}\sum_{n=1}^{N}\left|c_{g}\sum_{m=0}^{N}\mathcal{C}_{m,N-m}\mathcal{F}_{m,n}^{-g(N)}(\wp,\phi,\Lambda)-c_{e}\sum_{m=1}^{N}\mathcal{C}_{m-1,N-m}\mathcal{F}_{m,n}^{-e(N)}(\wp,\phi,\Lambda)\right|^{2},\label{w1}
\end{eqnarray}
where we define the two-dimensional Fourier transforms 
\begin{equation}
\mathcal{F}_{m,n}^{\pm\varepsilon(N)}(\wp,\phi,\Lambda)	=\int_{0}^{\infty}\int_{0}^{2\pi}\frac{d\theta d\rho}{2\pi k}\rho f\left(\frac{\rho}{k},\theta\right)D_{m-\delta_{\varepsilon e},n-\delta_{\varepsilon e}}^{(N-\delta_{\varepsilon e})}\left(\theta\right)\operatorname*{e}\nolimits ^{-i\rho\left[\wp\cos(\theta-\phi)\mp\sqrt{n}\Lambda\right]},\label{f1}
\end{equation}
with $\rho=kr$, $\varepsilon$ standing for the atomic states $g$ or $e$, $\delta_{\varepsilon,e}$ for the Kronecker delta ($\delta_{e,e}=1$,$\delta_{g,e}=0$). 

In order to generate the 2D momentum distribution, we have to analytically compute the Fourier integrals in Eq.\eqref{f1}. To this end, we assume the exponential azimuthal spatial distribution of the atoms
\begin{equation}	
f\left(\frac{\rho}{k},\theta\right)=\frac{1}{\sqrt{2\pi}\Delta r}\exp\left(-\frac{\rho}{2k\Delta r}\right). \label{density}
\end{equation}
 Inserting Eq.\eqref{density} into Eq.\eqref{f1} we obtain
\begin{equation}	
\mathcal{F}_{m,n}^{\pm\varepsilon(N)}\left(\wp,\phi,\Lambda\right)=\sum_{\ell=\max(0,m+n-N-\delta_{\varepsilon e})}^{\min(m-\delta_{\varepsilon e},n-\delta_{\varepsilon e})}\sum_{s=0}^{N-u+\delta_{\varepsilon e}}\sum_{t=0}^{u-2\delta_{\varepsilon e}}\left(ie^{i\phi}\right)^{v+\delta_{\varepsilon e}}\mathcal{R}_{m,n,\ell,s,t}^{\varepsilon(N)}\mathcal{S}_{n,s,t}^{\pm\varepsilon\left(N\right)}\left(\wp,\Lambda\right),\label{f2}
\end{equation}
where
\begin{equation}
\mathcal{R}_{m,n,\ell,s,t}^{\varepsilon(N)}=\frac{\left(-1\right)^{u-t-2\delta_{\varepsilon e}}}{2^{N-\delta_{\varepsilon e}}i^{u-2\delta_{\varepsilon e}}}\left(\begin{array}{c}
N-u+\delta_{\varepsilon e}\\s
\end{array}\right)\left(\begin{array}{c}
u-2\delta_{\varepsilon e}\\t
\end{array}\right)\overline{D}_{m-\delta_{\varepsilon e},n-\delta_{\varepsilon e},\ell}^{(N-\delta_{\varepsilon e})}
\end{equation}
\begin{equation}
\mathcal{S}_{s,t,n}^{\pm\varepsilon(N)}\left(\wp,\Lambda\right)=\frac{\left(-1\right)^{\Upsilon\left(v+\delta_{\varepsilon e}\right)}}{\sqrt{2\pi}k\Delta r}\frac{\left(\wp^{2}+\gamma_{\pm}^{2}\right)^{1/2}\left\vert v+\delta_{\varepsilon e}\right\vert +\gamma_{\pm}}{\left(\wp^{2}+\gamma_{\pm}^{2}\right)^{3/2}}\left(\frac{\wp}{\gamma_{\pm}+\left(\gamma_{\pm}^{2}+\wp^{2}\right)^{1/2}}\right)^{\left\vert v+\delta_{\varepsilon e}\right\vert },
\end{equation}
with $u=m+n-2\ell$, $v=2\left(s+t\right)-N$, and
\begin{eqnarray}
\Upsilon(\tilde{\nu})=\left\{ \begin{array}{ccc}
0 & \text{for even/odd} & \tilde{\nu}\geq0\\
0 & \text{for even} & \tilde{\nu}<0\\
1 & \text{for odd} & \tilde{\nu}<0
\end{array}\right.,\\ &&
\gamma_{\pm}\left(\Lambda\right)=-\left(2k\Delta r\right)^{-1}\pm i\sqrt{n}\Lambda.
\end{eqnarray}

Therefore, from the analytical expressions for the Fourier transforms given by Eq.\eqref{f2}, we readily obtain the atomic momentum distribution \eqref{w1}. The expressions in the main text are obtained by simply setting $c_{g}=0$ and $c_{e}=1$. Observe that for a finite number of total excitation $N$, the sums are finite so $W$ is actually composed of a finite number of terms.

\end{document}